\documentclass[final,1p,times]{elsarticle} 
\usepackage{graphicx} 
\usepackage{amssymb} 
\usepackage{amsthm} 
\usepackage{lineno} 
\usepackage{subfigure}
\usepackage[small,bf]{caption}
\usepackage{amsmath}

\journal{Nuclear Physics A} 
\begin{document} 

\begin{frontmatter} 


\title{Open Heavy Flavor Production at Forward Rapidity in $\sqrt{s_{NN}} =
200$ GeV Cu+Cu Collisions}

\author{Irakli Garishvili$^{a}$ for the PHENIX collaboration}

\address[a]{University of Tennessee, 
217 Nielsen Physics Building,
Knoxville, TN, 37996, USA}

\begin{abstract} 
The first measurement of the nuclear modification factor, $R_{AA}$, for single muons produced in Cu+Cu collisions at $\sqrt{s_{NN}} = 200$ GeV is presented, which is the first measurement of the nuclear modification factor for open heavy flavor production made at forward angles in heavy ion collisions.    Single muons are used to tag heavy flavor quark production via semi-leptonic decay of open heavy flavor mesons. With its muon arms, PHENIX has a unique ability to measure single muons at forward rapidity
($1.2 < \vert\eta\vert < 2.2$). 
\end{abstract} 

\end{frontmatter} 



\section{Single muon production in Cu+Cu collisions}

\begin{figure}[p]
\begin{minipage}[t]{0.45\linewidth} 
\hspace{-0.2cm}
\includegraphics[scale=0.35,angle=90]{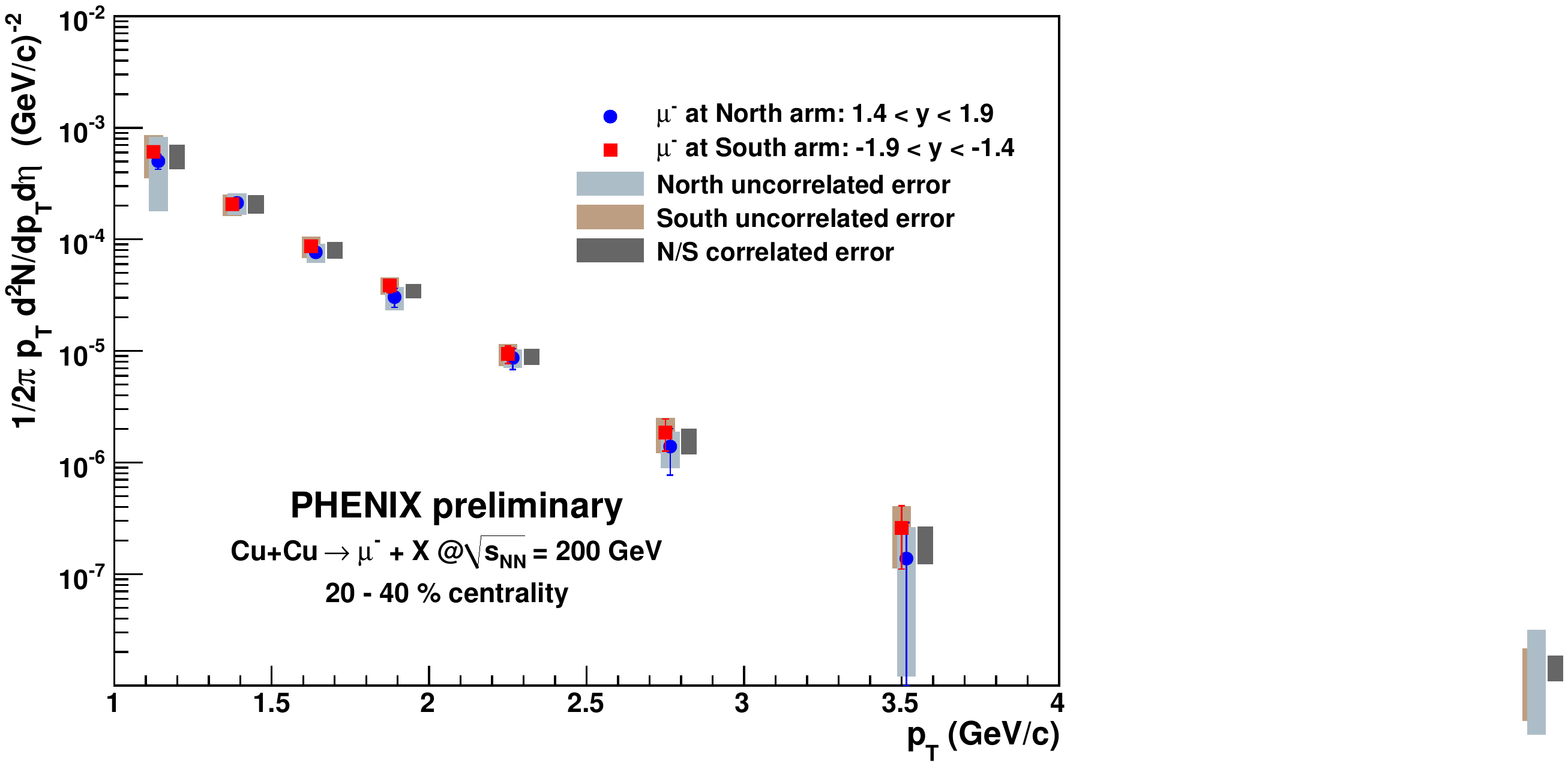}
\caption{The single muon invariant yield in mid-central (20-40\%) Cu+Cu collisions at
$\sqrt{s_{NN}}$ = 200 GeV separately measured in both muon arms.}
\label{both_arms}
\end{minipage}
\hspace{0.3cm} 
\begin{minipage}[t]{0.55\linewidth}
\centering
\includegraphics[scale=0.29]{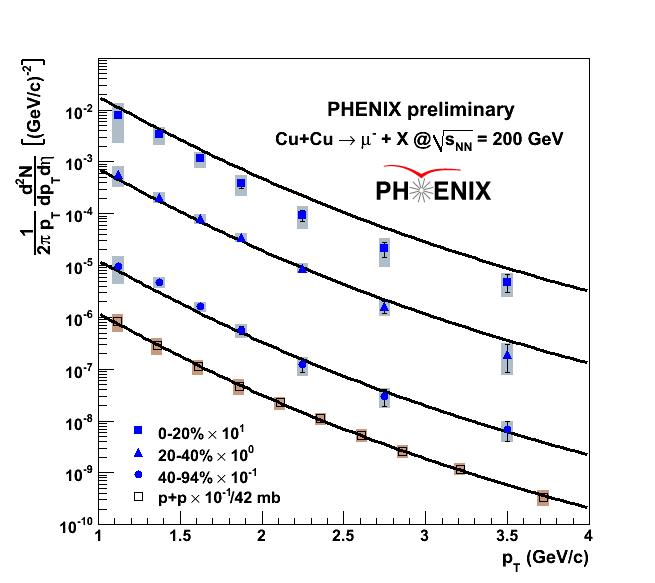}
\caption{Invariant yields of muons from heavy flavor decays for different Cu+Cu centrality classes (0-20\%, 20-40\% \& 40-94\%) and for p+p collisions \cite{donny_proc}, scaled by powers of ten for clarity. The solid lines are the result of a Kaplan function ($f(x)=p_{0}\left(1+\frac{x^{2}}{p_{1}}\right)^{p_{2}}$) fit to the p+p data, scaled with $\langle T_{AA}\rangle$ for each Cu+Cu centrality class.}
\label{all_spectra}
\end{minipage}
\vspace{2.0cm}
\begin{minipage}[b]{0.5\linewidth} 
\hspace{-0.2cm}
\includegraphics[scale=0.3]{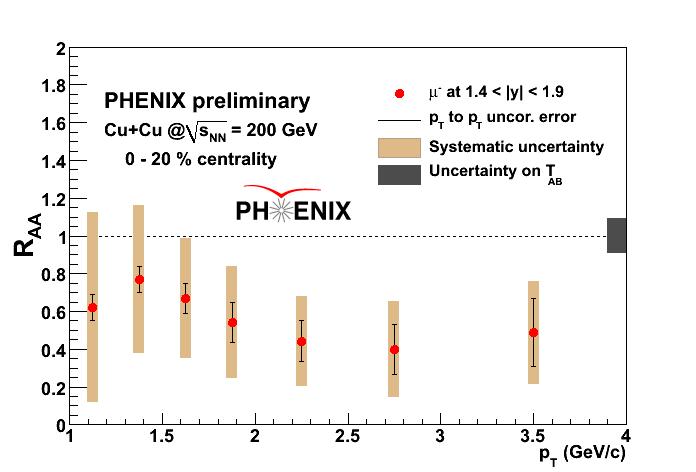}
\caption{The nuclear modification factor of the heavy flavor muons measured at $1.4<\vert y\vert <1.9$ as a function of $p_{T}$ in the most central (0-20\%) Cu+Cu collisions at $\sqrt{s_{NN}}$ = 200 GeV.}
\label{muon_raa}
\end{minipage}
\hspace{0.2cm} 
\begin{minipage}[b]{0.5\linewidth}
\centering	
\includegraphics[scale=0.375,angle=90]{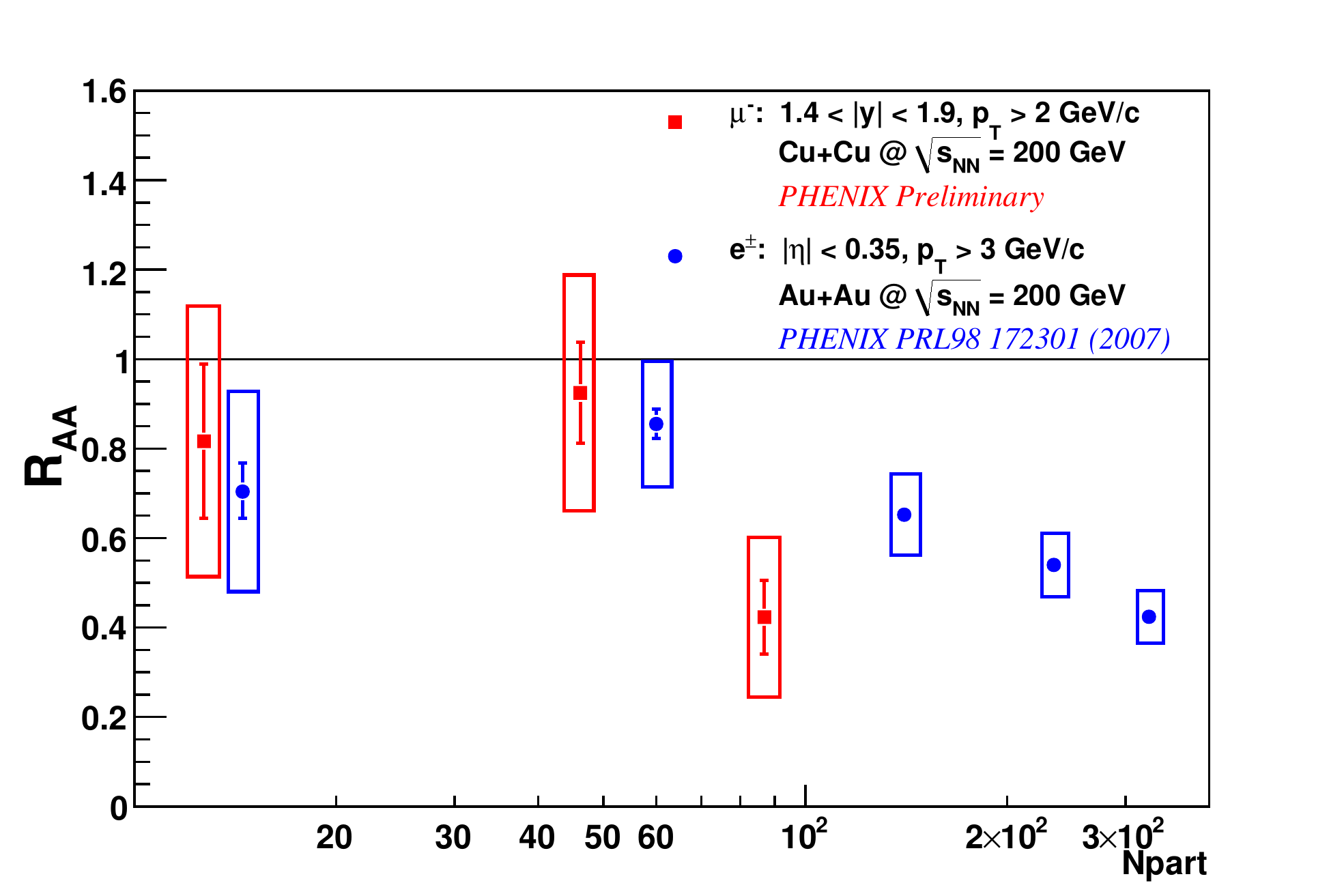}
\caption{$R_{AA}$ vs. Npart measured for single muons for $p_{T}>2$ GeV/c in Cu+Cu collisions at forward rapidity and non-photonic e$^{\pm}$'s for $p_T>3$ GeV/c in Au+Au collisions at mid-rapidity \cite{phe_elec_auau}.}
\label{el_vs_mu}
\end{minipage}
\end{figure}

The single muon signal is extracted via background subtraction from the raw spectra of muon candidate tracks. To predict the full background ``hadron cocktail", a full scale data-constrained Monte-Carlo simulation has been used. Realistic spectra of the dominant background sources ($\pi$, K, p, etc.) were generated as an initial estimation. The generated particles were propagated through the simulated geometry using GEANT. In the case where one is analyzing heavy-ion collisions, simulated tracks were ``embedded" into real events to reproduce the effects of the detector environment during the heavy ion collisions. Finally, the input spectra of the hadron cocktail were modified to match the hadron fluxes measured in the shallower layers of the Muon Identifier. Simultaneously, the background prediction was required to correctly reproduce the vertex dependence of the yield caused by the muons originating from the decays of light hadrons near the collision point. Single muons are separately measured in each of the two muon arms, as illustrated in Figure~\ref{both_arms}, which are then combined to form the final measurement. For symmetric collisions, the consistency between the two measurements leads to a reduction in the overall systematic uncertainties and provides further confidence in the final results.

The single muon spectra have been measured for three different centrality classes and compared to the baseline measurement made in p+p collisions \cite{donny_proc}, as shown in Figure~\ref{all_spectra}. These are then used to construct the $R_{AA}$ variable for each centrality class. Despite sizable systematic uncertainties, a significant suppression in single muon production is observed for the most central Cu+Cu collisions, as illustrated in Figure~\ref{muon_raa}, which is comparable to the suppression observed for high-$p_{T}$ non-photonic electrons at mid-rapidity measured in the most central Au+Au collisions shown on Figure~\ref{el_vs_mu}.
 
\end{document}